# Preparation and ionic conductivity of new $Bi_{12.5}Lu_{1.5}ReO_{24.5}$ phase

A.N. Bryzgalova[1], N.I. Matskevich[1*], Th. Wolf[2], C. Greaves[3]

[1]*Nikolaev Institute of Inorganic Chemistry, Siberian Branch of the Russian Academy of Science, Novosibirsk, 630090, Russia*

[2]*Karlsruhe Institute of Technology, Institute of Solid State Physics, Karlsruhe, D-76344, Germany*

[3]*School of Chemistry, University of Birmingham, Birmingham, B15 2TT, United Kingdom*

## 1. Introduction

At present the stabilized bismuth sesquioxide is one of the promising materials to be used for ceramic oxygen generators, electrocatalic activity (for the interconversion of molecular $O_2$ and $O^{2-}$ ions) and as an electrolyte in full cells operated at lower temperatures (<1200 K) [1-10]. Pure bismuth sesquioxide has two thermodynamically stable crystallographic polymorphs. One is $\alpha$-$Bi_2O_3$, which is stable below 1000 K and has a monoclinic structure, which shows p-type conduction. The other is $\delta$-$Bi_2O_3$, which is stable above 1000 K up to its melting temperature of 1100 K and crystallizes in the fluorite (cubic, $CaF_2$) structure. The $CaF_2$-type $\delta$-$Bi_2O_3$ contains 25% of the anion sites (one oxygen site per formula) vacant, and as a result exhibits very high $O^{2-}$ ion conductivity. The conductivity is up to two orders of magnitude greater than that in stabilized zirconia. However, the high conductivity phase is stable over narrow range of temperature (1000-1100 K). Further, the volume change associated with the $\delta \rightarrow \alpha$ transition leads to cracking and severe deterioration of the materials. Thus, for application of $Bi_2O_3$ as a solid electrolyte in fuel cells, it is imperative that the high-temperature

---

[*]Corresponding author. Tel.: +7–383–3306449; fax: +7–383–3306449

E–mail address: nata@niic.nsc.ru

cubic phase is stabilized. A large number of studies has shown that the high conductivity δ - phase in $Bi_2O_3$ could be stabilized at lower temperature by the addition of dopants (by various di-, tri-, tetra-, penta-, or hexavalent cations).

At present ones of the best ionic conductors are materials of type BIMEVOX (ME = Cu, Fe, Cr, etc.) [3]. Their main disadvantages are low mechanical strength and easiness of vanadium reduction when high activity of hydrogen. From this point of view it seems to be appropriate to carry out the directed synthesis of solid electrolytes at lower temperatures with the following characteristics: 1) ionic conductivity close to BIMEVOX; 2) isotropic; 3) low thermal expansion coefficient; 4) without ions, which can oxidize or reduce in the operating regime of the fuel cells. The new compounds of general formula $Bi_{12.5}ReLnO_{24.5}$ (Ln = Y, Eu, Er, La, Nd) were discovered in 2006 by Prof. C. Greaves [1-2]. Their conductivity in the temperature range 600-900 K is practically the same like BIMEVOX but they are isotropic and have a low thermal expansion coefficient. In pioneer works of Prof. C. Greaves the structure and conductivity of compounds with Ln = Y, Eu, Er, La, Nd were investigated.

Here we report the synthesis of a new compound, $Bi_{12.5}Lu_{1.5}ReO_{24.5}$, its structure, and conductivity measurements.

## 2. Experimental section

In literature [1-2] $Bi_2O_3$, $NH_4ReO_4$, and $R_2O_3$ were used to synthesize the compounds of general formula $Bi_{12.5}ReLn_{1.5}O_{24.5}$ (Ln = Y, Eu, Nd, La, Er). The temperature, used for synthesis was 1070 K. The temperature was selected because the δ - $Bi_2O_3$ phase exists in the temperature range of 1000-1100 K.

We used the following compounds to prepare $Bi_{12.5}Lu_{1.5}ReO_{24.5}$: $Bi_2O_3$ (99.999%, ABCR, Karlsruhe), $Re_2O_7$ (99.99%, Alfa Aesar), $Lu_2O_3$ (99.999%, ChemPur). Synthesis proceeds according to the reaction:

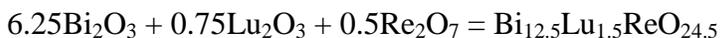

$6.25Bi_2O_3 + 0.75Lu_2O_3 + 0.5Re_2O_7 = Bi_{12.5}Lu_{1.5}ReO_{24.5}$

The samples were prepared as follows. Starting reagents were mixed in an agate mortar and ground for about 70 h with four intermediate reground in a planetary mill (FRITSCH pulverisettee, analyseette Laborrette). All operations with $Re_2O_7$ were performed in a glove box under pure argon, because $Re_2O_7$ reacts with water immediately. Then samples were placed in the furnace (Carbolite) and hold at 1070 K in air for 70 h. The same technique was used to prepare high quality samples of other complex oxides [11, 12].

The phase purity of the samples was analyzed with an X-ray diffractometer (STADI-P, Stoe diffractometer, Germany, Cu $K_{\alpha 1}$ radiation). Impurities were determined by X-ray Fluorescence Spectrometer (ARL ADVANT'XP). A typical X-ray Spectrum is presented in Fig. 1.

X-ray power diffraction indicated that the sample is single phase and that the δ - phase structure had been stabilized to room temperature.

Structural investigated were based on neutron powder diffraction data (collected at Studsvik, Sweden; 10 K; wavelength 1.4703 Å). A detailed description of structural determination is given in ref. [1-2]. Rietveld refinements were performed using the program GSAS with a starting model based on pure δ-$Bi_2O_3$, space group Fm-3m, with the cations statistically distributed on the $4a(0,0,0)$ site and the oxygen atoms occupying the regular $8c$ and interstitial $32f(x,x,x)$ site. The fitted profile for $Bi_{12.5}Lu_{1.5}ReO_{24.5}$ (see, Fig. 2) is typical and indicates good agreement between observed and calculated profiles; an undulating background reflects the high level of disorder. The refined parameters are shown in Table 1.

| Atom | X | y | Z | Ui/Ue*10 | Fractional occupancy |
|---|---|---|---|---|---|
| Bi | 0.0 | 0.0 | 0.0 | 4.54(3) | 0.8333 |
| Lu | 0.0 | 0.0 | 0.0 | 4.54(3) | 0.10 |
| Re | 0.0 | 0.0 | 0.0 | 4.54(3) | 0.0667 |
| O(1) | 0.25 | 0.25 | 0.25 | 11.5(1) | 0.593(5) |
| O(2) | 0.364(1) | 0.364(1) | 0.364(1) | 11.5(1) | 0.060(1) |

Table 1. Refined Structural data for $Bi_{12.5}Lu_{1.5}ReO_{24.5}$

As mentioned before, the $Bi_{12.5}ReLn_{1.5}O_{24.5}$ (Ln = Y, Eu, Nd, La, Er) compounds have been reported in literature. In reference [1] the lattice parameters were given as follows: a = 5.6456(3) Å ($Bi_{12.5}La_{1.5}ReO_{24.5}$), a = 5.6184(4) Å ($Bi_{12.5}Nd_{1.5}ReO_{24.5}$), a = 5.5689(5) Å ($Bi_{12.5}Er_{1.5}ReO_{24.5}$), a = 5.575(1) Å ($Bi_{12.5}Y_{1.5}ReO_{24.5}$). For $Bi_{12.5}Lu_{1.5}ReO_{24.5}$ we determined the lattice parameter a = 5.5591(2) Å. With the ionic radii for the rare-earth elements taken from Shannon [13] ($La^{+3}$ = 1.032 Å, $Nd^{+3}$ = 0.983 Å, $Y^{+3}$ = 0.900 Å, $Er^{+3}$ = 0.890 Å, $Lu^{+3}$ = 0.861 Å) it can be seen that the lattice parameters increased with increasing ionic radius. Fig. 3 shows that this increase is linear.

Impedance measurements were made on pellets (about 70% of theoretical density; gold contacts and wires) in the range 500-900 K using Hewlett-Packard 4192A and 4800A impedance analyzers. The conductivities (see, Fig. 4) were calculated from the overall resistance determined from the minima in the complex plane plots. In the article [1] the data of conductivity for $BiCuVO_x$, and $Bi_{12.5}Ln_{12.5}ReO_{24.5}$ (Ln = Eu, Er, La, Nd, Y) are presented. A comparison of the reported data with the conductivity of our sample shows that the $Bi_{12.5}Lu_{1.5}ReO_{24.5}$ phase appears to be one of the best moderate temperature isotropic ion conductors known. For example, the conductivity of $Bi_{12.5}Lu_{1.5}ReO_{24.5}$ at 800 K is equal to the conductivity of $BiCuVO_x$, $Bi_{12.5}Ln_{12.5}ReO_{24.5}$ (Ln = Eu, Er, La, Nd, Y) phases.

**Conclusions**

As a conclusion it is possible to say that new $Bi_{12.5}Lu_{1.5}ReO_{24.5}$ phase has been synthesized and shown high ion conductivity at the moderate temperatures. Structural analysis shows that space group is Fm3m with lattice parameter a = 5.5591(2) Å. The conductivity was measured in the temperature range of 600-800 K. The conductivity of $Bi_{12.5}Lu_{1.5}ReO_{24.5}$ at 800 K is the same as the conductivity of $BiCuVO_x$, $Bi_{12.5}Ln_{12.5}ReO_{24.5}$ (Ln = Eu, La, Nd) phases. In this connection the $Bi_{12.5}Lu_{1.5}ReO_{24.5}$ phase offers excellent potential for moderate temperature application.

**Acknowledges**

This work is supported by Karlsruhe Institute of Technology, DFG (grant LO 250/24-1), NATO programme (grant CBR.NR.NRCLG 982559) and Program of Fundamental Investigation of Siberian Branch of the Russian Academy of Sciences.

**Abstract**

The substitution of Re into $Bi_2O_3$ allows stabilization of the δ- $Bi_2O_3$ structure by additional substitution of lutetium ion to give phase of composition $Bi_{12.5}Lu_{1.5}ReO_{24.5}$. The phase was synthesized for the first time. Structural analysis performed by neutron diffraction showed that space group was Fm3m with lattice parameter a = 5.5591(2) Å. The phase has been found to show high ion conductivity at moderate temperature. The conductivity was measured in the temperature range of 600-800 K. The conductivity of $Bi_{12.5}Lu_{1.5}ReO_{24.5}$ at 800 K is the same as the conductivity of $BiCuVO_x$, $Bi_{12.5}Ln_{1.5}ReO_{24.5}$ (Ln = Eu, La, Nd) phases. In this connection the $Bi_{12.5}Lu_{1.5}ReO_{24.5}$ phase offers excellent potential for moderate temperature application.

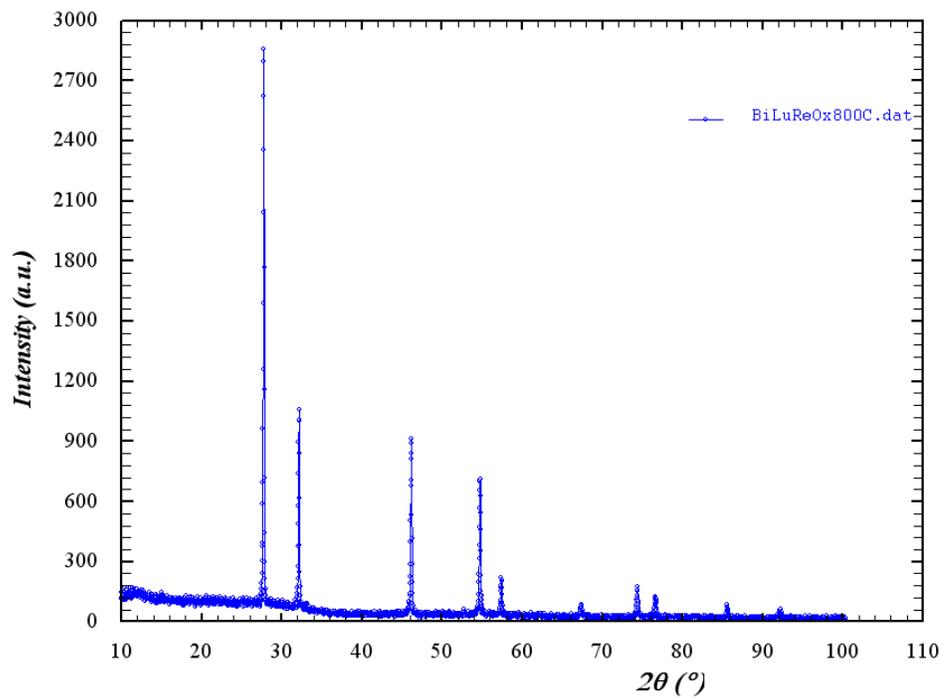

Fig.1. X-ray diffraction of $Bi_{12.5}Lu_{1.5}ReO_{24.5}$ at room temperature

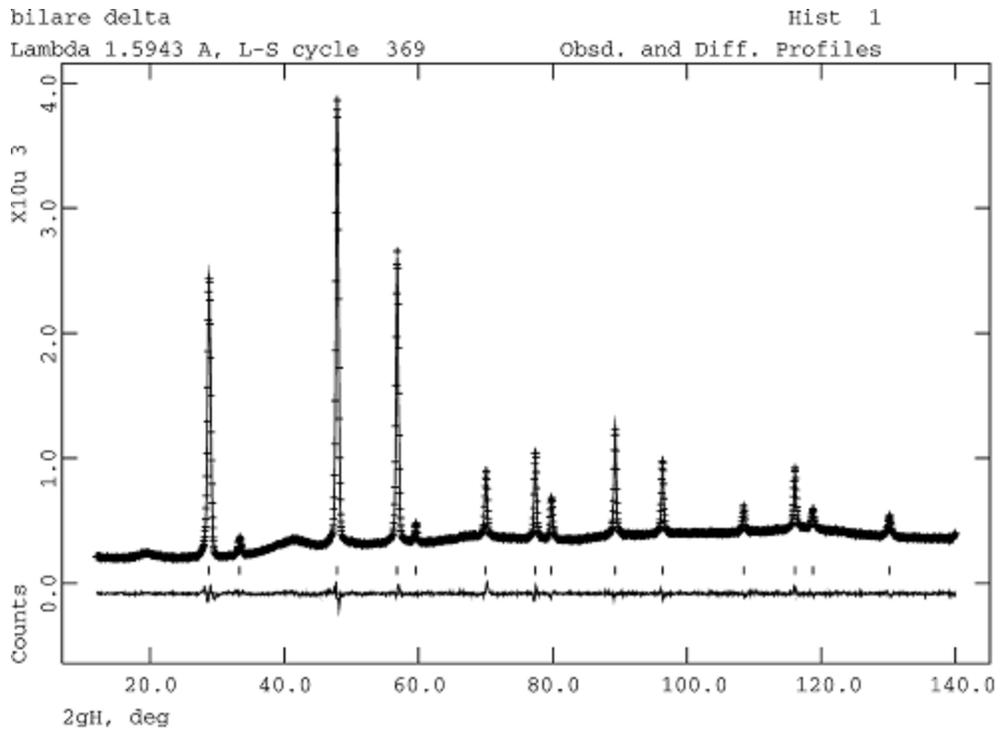

Fig. 2. Neutron diffraction data for $Bi_{12.5}ReLu_{1.5}O_{24.5}$ at 10 K

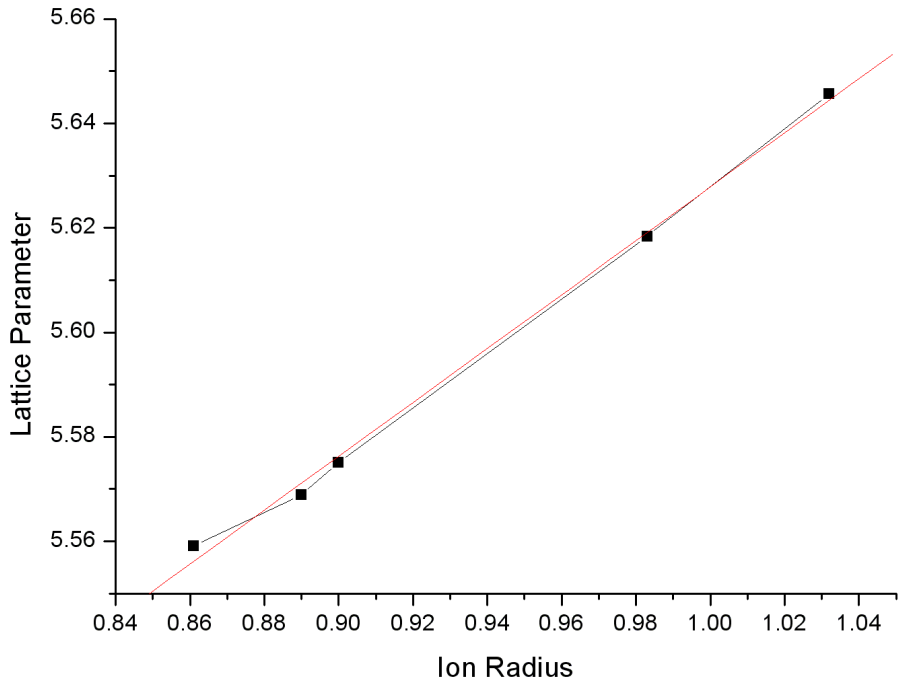

Fig. 3. Dependence of lattice parameters from ionic radii

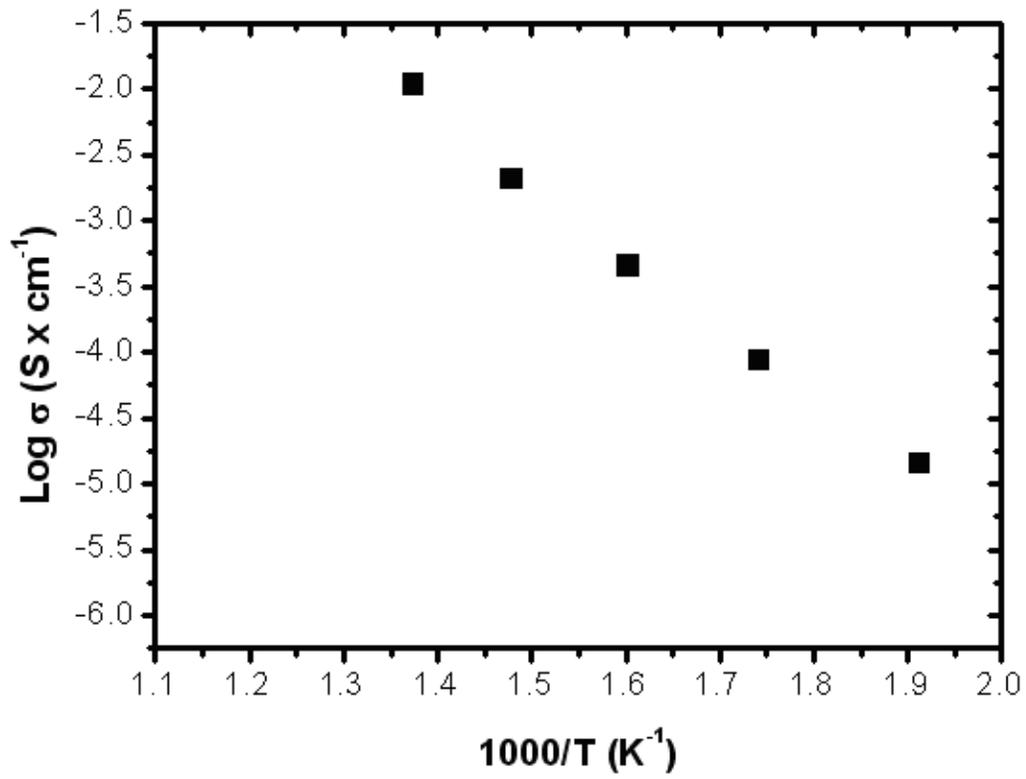

Fig. 4. Variation of conductivity with temperature for $Bi_{12.5}Lu_{1.5}ReO_{24.5}$